# aIR-Jumper:
# Covert Air-Gap Exfiltration/Infiltration via Security Cameras & Infrared (IR)

Mordechai Guri,[1] Dima Bykhovsky[2], Yuval Elovici[1]
[1]Cyber Security Research Center, Ben-Gurion University of the Negev, Israel
[2] Electrical and Electronics Engineering Department, Shamoon College of Engineering, Israel
gurim@post.bgu.ac.il; dmitrby@ac.sce.ac.il; elovici@bgu.ac.il

demo video: http://cyber.bgu.ac.il/advanced-cyber/airgap

*Abstract*— Infrared (IR) light is invisible to humans, but cameras are optically sensitive to this type of light.

In this paper, we show how attackers can use surveillance cameras and infrared light to establish bi-directional covert communication between the internal networks of organizations and remote attackers. We present two scenarios: exfiltration (leaking data out of the network) and infiltration (sending data into the network). **Exfiltration.** Surveillance and security cameras are equipped with IR LEDs, which are used for night vision. In the exfiltration scenario, malware within the organization access the surveillance cameras across the local network and controls the IR illumination. Sensitive data such as PIN codes, passwords, and encryption keys are then modulated, encoded, and transmitted over the IR signals. An attacker in a public area (e.g., in the street) with a line of sight to the surveillance camera records the IR signals and decodes the leaked information. **Infiltration.** In an infiltration scenario, an attacker standing in a public area (e.g., in the street) uses IR LEDs to transmit hidden signals to the surveillance camera(s). Binary data such as command and control (C&C) and beacon messages are encoded on top of the IR signals. The signals hidden in the video stream are then intercepted and decoded by the malware residing in the network. The exfiltration and infiltration can be combined to establish bidirectional, 'air-gap' communication between the compromised network and the attacker. We discuss related work and provide scientific background about this optical channel. We implement a malware prototype and present data modulation schemas and a basic transmission protocol. Our evaluation of the covert channel shows that data can be covertly exfiltrated from an organization at a rate of 20 bit/sec per surveillance camera to a distance of tens of meters away. Data can be covertly infiltrated into an organization at a rate of over 100 bit/sec per surveillance camera from a distance of hundreds of meters to kilometers away. These transmission rates can be increased further when several surveillance cameras are used.

## I. INTRODUCTION

Organizations often protect their internal networks from Internet attacks by using firewalls, intrusion detection systems (IDSs) and intrusion prevention systems (IPSs). For a higher degree of protection, so-called 'air-gap' isolation is used. In this case, the internal network is kept disconnected from the Internet, without any physical or logical connection. Air-gap isolation is commonly used in military networks, critical infrastructures, financial industries, and other settings [1] [2].

In the past decade it has been shown that firewalls, IDSs, IPSs and air-gap isolation do not provide hermetic protection. Motivated attackers can evade any level of separation and infect an organization's internal network with malware. In order to breach the target networks, attackers have used complex attack vectors, such as supply chain attacks, malicious insiders, and social engineering [3].

In 2008, a classified network of the United States military was compromised by a computer worm named Agent.Btz [3]. According to reports [4], a foreign intelligence agency supplied infected thumb drives to retail kiosks near NATO headquarters in Kabul. A malicious thumb drive was put into a USB port of a laptop computer that was attached to United States Central Command. The worm spread further to both classified and unclassified networks. Other attacks on secure networks in the governmental, financial, and IT sectors have also been reported [5] [6] [7] [8] [9].

### A. Covert Channels

With its malware deployed on the target network, the attacker might wish to establish a communication with it. For example, the attacker may want to deliver command and control (C&C) messages to the malware or leak data out of the compromised network. Over the years, various covert communication channels have been proposed. These communication channels allow attackers to communicate with highly secured networks while bypassing firewalls, IDSs, and IPSs. To evade detection, attackers may hide the data within TCP/IP headers, HTTPS requests, transferred images, and other data streams sent over the network. However, the traditional covert channels depend on network connectivity between the attacker and the target network.

### B. *Air-Gap Covert Channels*

Air-Gap covert channels are a special type of covert channel that can operate on air-gapped networks (i.e., they are a covert channel that doesn't require network connectivity between the attacker and the target network). In air-gap covert channels, attackers may leak data through different types of radiation emitted from the computer. Leaking data using electromagnetic radiation has been investigated for more than twenty years. In this method, malware controls the electromagnetic radiation emitted from computer parts, such as LCD screens, communication cables, computer buses, and other components [10] [11] [12] [13] [14]. Other air-gap covert channels are based on sonic and ultrasonic sound [15] [16] and heat emissions [17]. In optical covert channels, information is leaked via optical signals controlled by the malware. The optical signals can be generated by the hard drive activity LEDs [18], keyboard LEDs [19], router LEDs [20], implanted IR LEDs [21], and via invisible images on the screen [22]. Most optical methods are not considered completely covert, since they can easily be detected by people who notice the optical activity (e.g., blinking LEDs).

### C. *Surveilance and Security Cameras*

In recent years, there were several cases in which security and surveillance cameras has been compromised by attackers [23] [24]. For example, in January 2017, two hackers were reportedly arrested in London on suspicion of hacking 70% of the CCTV cameras in Washington [25]. A comprehensive analysis of the threats, vulnerabilities, and attacks on video surveillance, closed-circuit TV, and IP camera systems is presented by Costin in [26].

### D. *Our Contribution*

In this paper we introduce a new type of covert channel that uses infrared (IR) light and surveillance cameras for data exfiltration and infiltration.

**Exfiltration.** Many surveillance and security cameras are equipped with IR LEDs which enable night vision. We show that malware residing within the internal networks of the organization can control these IR LEDs, turning them on and off or controlling their IR intensity. We implement a malware prototype and show that binary data can be encoded over the IR signals and leaked to an attacker from a distance of tens of meters away. Notably, many surveillance and security cameras monitor public areas, and therefore attackers can easily establish a line of sight with them.

**Infiltration.** Cameras can 'see' light at the IR wavelength but humans cannot. We show that an attacker can generate IR signals that will be recorded by the surveillance cameras, in order to deliver C&C messages to malware within the organization. With access to the video stream recorded by the surveillance camera, the malware can detect the covert signals and decode the C&C messages.

The proposed covert channel has the following advantages: (1) **invisibility**: IR is invisible to humans, hence our method is considered optically covert. Most existing optical channels use light in the visible range (e.g., computer LEDs) which is visible to humans, (2) **software based:** our method doesn't require dedicated hardware and simply exploits surveillance cameras that exist in all secured organizations today, (3) **communication over the air-gap**: our method works even when there is no network connection between the attacker and the malware in the organization.

The remainder of the paper is organized as follows. Section II describes the attack scenarios. In Section III we present related work. Technical background about IR light and cameras is provided in Section IV. Data encoding and decoding are discussed in Section V. Signal generation is described in Section VI. In Section VII we present the evaluation and analysis. Countermeasures are discussed in Section VIII, and we present our conclusions in Section IX.

## II. ATTACK SCENARIOS

We introduce an IR based covert channel which is relevant to two scenarios; exfiltration and infiltration. In the exfiltration scenario, information is leaked out from the organization's internal networks to a remote attacker. Such information includes passwords, PIN codes, encryption keys, and keylogging data. In the infiltration scenario, information is delivered from a remote attacker to the organization's internal networks. Such information might consist of C&C messages for malware residing in the network.

### A. *Intrusion & Infection*

Like a typical advanced persistent threat (APT), in the initial stage of the attack an internal network within the organization is compromised with malware. Compromising secured and even isolated (air-gapped) networks can be accomplished with various attack vectors such as supply chain attacks, malicious insiders, or social engineering [3]. Once within the network, the malware scans the IPs to find the surveillance and security cameras connected to the local network. Surveillance cameras can be located in the network by examining the open ports, HTTP response, or MAC access of the objects in the network. After mapping the surveillance cameras the malware must be able to connect to the cameras in order to receive their input stream and modify their settings (e.g., control the IR LEDs). Connecting to the camera usually requires a password which can be retrieved from a computer in the network that has an access to the cameras, or by exploiting a vulnerability in the camera's software/firmware [26] [25] [23] [24].

### B. *Exfiltration*

The exfiltration scenario is illustrated in Figure 1. The malware in the network collects sensitive data that it wants to exfiltrate. When the data is collected, the malware transmits it by encoding it over the IR signals emitted from the camera's night vision IR LEDs. Exfiltration may take place at predefined times or as the result of a trigger from the attacker side. An attacker located outside the secured facility (e.g., on the street) can receive the IR signals by carrying a standard video camera that is aimed at the transmitting surveillance camera. The received video is then processed in order to decode the transmitted data.

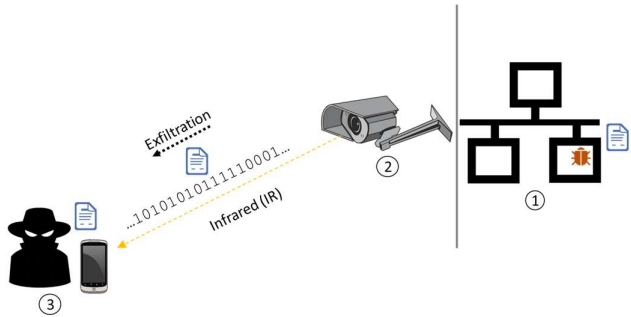

**Figure 1. Illustration of the exfiltration covert channel; malware accesses the surveillance camera and transmits invisible signals to a remote attacker by using the IR LEDs on cameras.**

### C. *Infiltration*

The infiltration scenario is illustrated in Figure 2. An attacker located outside the secured facility (e.g., on the street) generates invisible IR signals by using IR LEDs. The IR signals are modulated with the C&C messages to be delivered to the malware. The video stream recorded by the surveillance camera is received by the malware which processes and decodes the transmitted data.

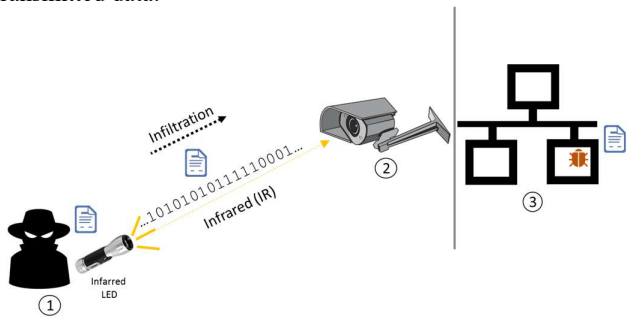

**Figure 2. Illustration of the infiltration covert channel; IR LEDs transmit invisible signals to the camera. The signals are received by malware with access to the surveillance camera.**

Table 1 provides some details regarding the two scenarios.

**Table 1. The exfiltration and infiltration scenarios**

| Scenario | Transmitter | Receiver | Type of transmissions |
|---|---|---|---|
| **Exfiltration (leaking data out of the network)** | Surveillance cameras' IR LEDs | Video camera | Brief information (e.g., PIN codes, passwords, encryption keys). |
| **Infiltration (sending C&C messages to the network)** | IR LEDs | Surveillance cameras | Command and Control (C&C) messages |

### D. *Using Several Cameras*

Organizations often use CCTV systems consisting of many outdoor surveillance cameras to secure their facilities and monitor the areas around them. Within the context of this covert channel, this means that the attacker can choose the best location to conduct the attack. In addition, the covert channel can be established with more than one surveillance camera in order to multiply the channel's bandwidth. Figure 3 presents the layout of surveillance cameras in a typical secured building. Each of the three outdoor cameras can exfiltrate data to an attacker with a line of sight to the camera, and each of these cameras can receive IR signals from a remote attacker.

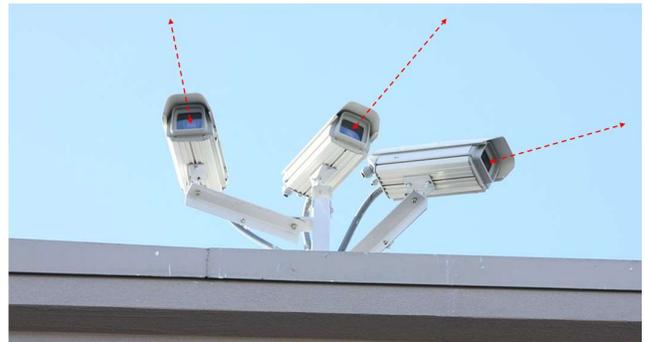

**Figure 3. The layout of three surveillance cameras in a typical secured building. Each of the cameras can be used for exfiltration and infiltration from different angles and positions.**

### E. *Other Types of Cameras*

Besides indoor and outdoor surveillance cameras, many doorbell cameras are also equipped with controllable IR LEDs (Figure 4). Like surveillance cameras, these types of cameras can also be used for exfiltration and infiltration. Note that doorbell cameras are typically installed at locations and heights which make it easy for an attacker to obtain a line of sight. In a typical scenario the attacker stands in front of, or at a short distance from, a doorbell camera while exchanging data via covert IR signals.

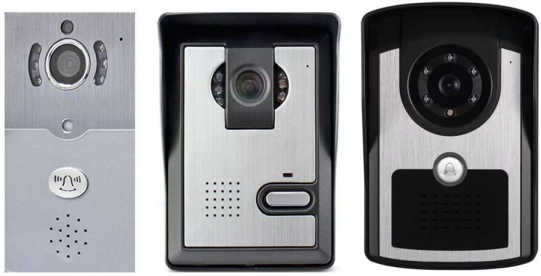

**Figure 4. Doorbell cameras with IR LEDs**

### III. RELATED WORK

Conventional covert communication channels depend on network connectivity between the attacker and the target network. In these covert channels, attackers may hide the data within TCP/IP headers, HTTPS requests, transferred images, and other data streams sent over the network [27] [28]. However, in cases where there is no direct connection with the target network, the attacker may resort to so-called air-gap covert channels.

Communication over air-gaps can be established via electromagnetic, acoustic, thermal, and optical emanations. In 1998, Anderson [11] show how to leak data using electromagnetic waves emanating from display screens. AirHopper [10] [29] is malware introduced recently that exfiltrates data by using FM radio signals emanating from the video card. In 2015, Guri et al presented GSMem [14], malware that leaks data from air-gapped computers using cellular frequencies emitted from RAM buses. USB buses and GPIOs also generate electromagnetic signals that can be exploited for data exfiltration [30] [31].

Hanspach addresses covert channels via ultrasonic sound waves [32]. He used laptop speakers and microphones to transmit data between air-gapped computers. In 2016, Guri et al presented Fansmitter [33] and DiskFiltration [34], two methods enabling exfiltration of data via sound waves, when the computers are not equipped with speakers or audio hardware. They used the computer fans and hard disk drive to generate acoustic signals. BitWhisper [35], a method presented in 2015, established covert communication between two adjacent air-gapped computers using heat emissions.

Several optical channels have also been proposed over the years. Loughry studied data exfiltration via the keyboard LEDs [19]. In this attack, malware sends data by blinking the caps-lock, num-lock and scroll-lock LEDs. More recently, Guri et al presented a covert channel that enables data leakage from air-gapped networks via the hard drive indicator LED [18]. VisiSploit [22] is an optical based covert channel in which data is leaked through a hidden image projected on an LCD screen. The 'invisible' image is obtained by a remote camera and is then reconstructed using image processing. Lopes [21] presented an IR covert channel based on a malicious hardware component with implanted IR LEDs. By blinking the IR LEDs an attacker can leak sensitive data stored on the device, such as credentials and cryptographic keys, at a speed of 15 bit/sec. However, in their method the attacker must find a way to insert the compromised hardware into the organization. In contrast, our method uses the IR LEDs that already exist in surveillance and security cameras and doesn't require special or malicious hardware. Note that the threat of using the surveillance camera for exfiltration was also discussed by Costin in its comprehensive analysis on the security of CCTV system [26]. Our paper focus on this threat and presents the attack model, design, analysis, implementation and evaluation of the attack.

A list of existing air-gap covert channels is presented in Table 2.

**Table 2. Different types of air-gap covert channels**

| Type | Method |
|---|---|
| Electromagnetic | AirHopper [11] [25] |
| | GSMem [15] |
| | USBee [26] |
| | Funthenna [13] |
| Acoustic | [17] [16] [30] [62] [53] |
| | Fan noise (Fansmitter) [32] |
| | Hard disk noise (DiskFiltration) [34] |
| Thermal | BitWhisper [34] |
| Optical | Hard Drive LED (LED-it-GO) [18] |
| | VisiSploit (invisible pixels) |
| | Keyboard LEDs [20] |
| | Screen LEDs [21] |
| | Implanted infrared LEDs [21] |
| | Router LEDs [16] |

Our optical covert channel has two unique characteristics:

**Invisibility.** Unlike existing optical covert channels which use visible light (e.g., keyboard LEDs), the proposed covert channel is based on IR light which is optically invisible to humans. That makes the channel more covert compared to the existing covert channels.

**Bidirectional.** The proposed channel can be used for both exfiltration and infiltration, which allows the attacker to establish bidirectional communication with the target network. Existing optical covert channels are only capable of exfiltration.

### IV. INFRARED (IR) AND CAMERAS

In this section, we briefly present the technical background relevant to the covert channel. We discuss IR light, the cameras and their sensitivity to IR light, and some optical aspects of the communication.

#### A. *Infrared Light*

A healthy human eye of an adult is sensitive to only a small segment of the optical wavelengths that exist, namely from

~400nm to ~700nm. This range is also known as the *visible range*. For a typical consumer camera, the sensitivity range is slightly wider, from ~350nm to ~1050nm, including a portion of near-ultraviolet (NUV) and near-infrared (NIR) light, as presented in Figure 5.

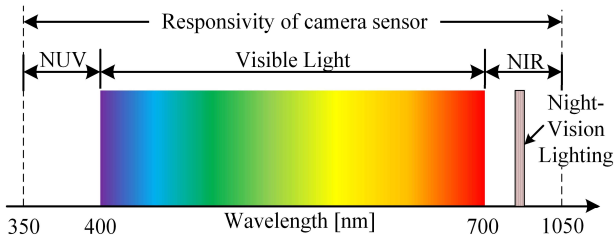

**Figure 5. A typical spectral sensitivity range of cameras.**

Typically, modern cameras (both video and still cameras) can record lighting in the NIR range. Figure 6 contains two versions of an IR projector that consists of 192 IR LEDs. Version 1 shows how the projector appears to the human observer, and version 2 shows how the IR LEDs on the projector appear in a photo taken by a Samsung Galaxy smartphone, thus demonstrating the ability of a camera to record lighting in the IR range.

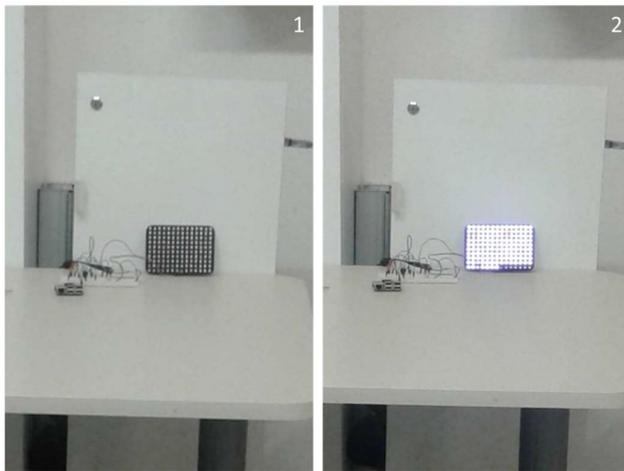

**Figure 6. An IR projector as it is appears to a human observer (1) and to a camera (in a photo taken by a smartphone camera) (2). The IR light is invisible to a human observer.**

### B. *Camera IR LEDs (night vision capability)*

In order to provide night vision capability, surveillance cameras use lighting in the NIR range. Note that the NUV range is not used due to the strict eye safety guidelines regarding the use of NUV light. Note that during the daytime, the presence/absence of NIR lighting has a negligible influence, since it has significantly less power than that of other typical types of outdoor illumination. Cameras with night vision capability have IR LEDs positioned around the outer edges of the camera lens which illuminate during low light conditions. With IR illumination cameras can capture objects from tens of meters away, even in total darkness. The camera's IR lights typically has central wavelength of about 850 nm, spectral full-width half-maximum (FWHM) of about 45-50 nm and power of few watts. They are also organized in arrays of LEDs for higher joint lighting power.

#### 1) IR LED Control

In many high-end surveillance cameras, the IR intensity levels can be adjusted to fit different environmental lighting conditions. Typically, low levels of IR illumination are used when there is ambient light in the area, whereas high levels are used in the case of total darkness or low ambient lightening. Figure 7 shows two high-end security cameras: the Sony SNC-EM602RC (top) and Sony SNC-EB600 (bottom), with their IR LEDs turned on. The intensity of IR increases from level 0 in the leftmost images to level 4 in the rightmost images.

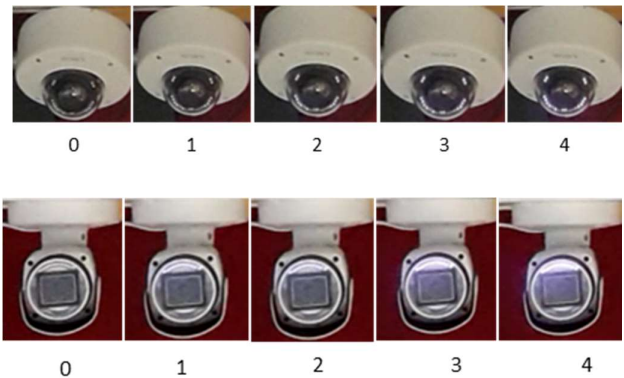

**Figure 7. Two high-end surveillance cameras, with five levels of IR intensity (0=lowest, 4=highest).**

As presented in this paper, the changes in the lighting level may be used for data modulation. The modulation frequency is limited only by the camera's hardware, since the maximum modulation frequency of the typical LEDs is about 2 MHz. The changes in lighting levels that are perceptible to a camera user are termed *flickering*. Nevertheless, if lighting change frequency is carefully chosen, flickering may be avoided [36].

### C. *Line-of-sight (LOS) and Non line-of-sight (NLOS)*

There are two ways of communicating between the camera and the IR source: line of sight communication (LOS) and non-line of sight (NLOS) commination. LOS communication is illustrated in Figure 8. For the exfiltration scenario (a), the IR LEDs of the surveillance camera are directed toward the NIR optical receiver. For the infiltration scenario (b), the NIR transmitter is directed toward the surveillance camera. NLOS communication is illustrated in Figure 9. For the exfiltration scenario (a), the area illuminated by the IR LEDs of the surveillance camera is located within the field of view (FOV) of the dedicated optical receiver. For the infiltration scenario (b), the NIR transmitter is directed toward the FOV of the surveillance camera. Note that in NLOS communication the receiver itself may be remain invisible from the transmitting surveillance camera. More detailed discussion regarding the effectiveness of LOS and NLOS communication is presented in Section VII.

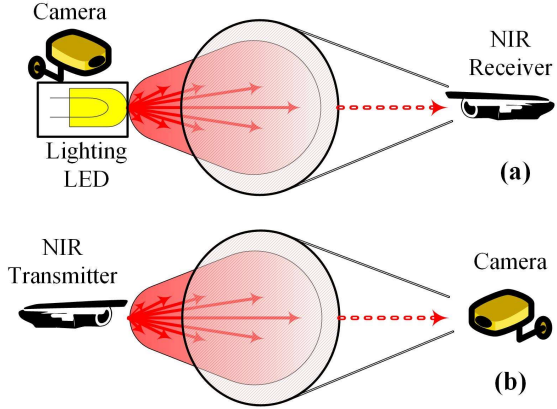

Figure 8. Illustration of LOS communication: (a) exfiltration and (b) infiltration.

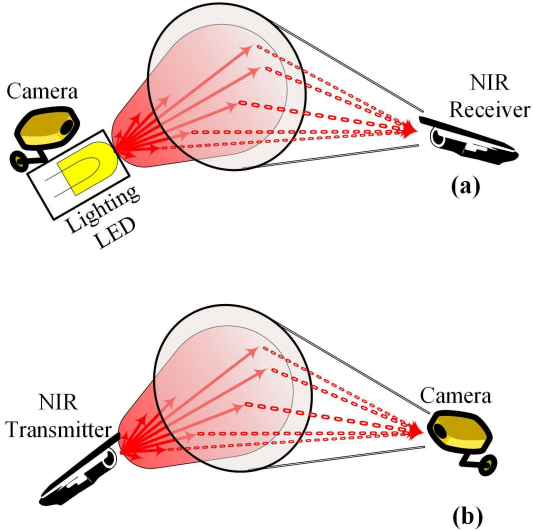

Figure 9. Illustration of NLOS communication: (a) exfiltration and (b) infiltration.

## V. Data Encoding & Decoding

In this section we discuss data encoding and decoding. Note that the topic of visible light communication (VLC) has been widely studied in the past. In particular, various modulations and encoding schemes have been proposed for LED communication [37] [38] [39]. We present basic encoding schemes and describe their characteristics, and also present simple bit framing. Recall that high-end cameras may have more than one level of IR illumination. Given $n$ possible levels, we denote them as LED-$ON_0$, LED-$ON_1$, LED-$ON_2$... LED-$ON_{n-1}$.

### A. Data Encoding

**On-off keying (OOK).** In this type of modulation, the absence of a signal for a certain duration encodes zero ('0'), while its presence for the same duration encodes one ('1'). In our case, LED-$ON_a$ for duration of $T_{off}$ encodes '0' and LED-$ON_b$ for a duration $T_{on}$ encodes '1,' where $0 \leq a,b < n-1$. The OOK encoding is described in Table 3.

Table 3. On-off-keying (OOK) modulation

| Logical bit | Duration | LED state |
|---|---|---|
| 0 | $T_{off}$ | LED-$ON_a$ |
| 1 | $T_{on}$ | LED-$ON_b$ |

**Frequency-shift keying (FSK).** In this modulation, the data is modulated through frequency changes in a carrier signal. In the simple form which is called binary frequency-shift keying (B-FSK), only two frequencies are used for modulation. In our case, LED-$ON_a$ for a duration of $T_{off}$ encodes '0' and LED-$ON_a$ for duration $T_{on}$ encodes '1,' where $0 \leq a,b < n-1$. We separate between two bits by setting the LED-$ON_b$ state for a time interval of $T_d$. The B-FSK encoding is described in Table 4.

Table 4. Binary frequency-shift keying (B-FSK) modulation

| Logical bit | Duration | LED state |
|---|---|---|
| 0 | $T_{off}$ | LED-$ON_a$ |
| 1 | $T_{on}$ | LED-$ON_a$ |
| Interval | $T_d$ | LED-$ON_b$ |

**Amplitude Shift Keying (ASK).** As was mentioned in the previous section, in professional surveillance cameras it is possible to control the level of the light illuminated by the IR LEDs. In ASK modulation, different levels (different amplitudes) of IR light are used to represent a different sequence of bits. Table 5 presents a case in which the IR LEDs can have five different levels: $A_0, A_1, A_2, A_3$ and $A_4$.

Table 5. Amplitude Shift Keying (ASK)

| Logical bits | Duration | Amplitude | LED state |
|---|---|---|---|
| 00 | $T_{on}$ | $A_1$ | LED-$ON_1$ |
| 01 | $T_{on}$ | $A_2$ | LED-$ON_2$ |
| 10 | $T_{on}$ | $A_3$ | LED-$ON_3$ |
| 11 | $T_{on}$ | $A_4$ | LED-$ON_4$ |
| Interval | $T_d$ | $A_0$ | LED-$ON_0$ |

In this case, setting the IR level to $A_1$ for a duration of $T_{on}$, encodes '00,' setting the IR level to $A_2$ for a duration of $T_{on}$ encodes '01,' and so on. We insert a separation between two sequential bytes by setting the LEDs in the $A_0$ level at a time interval of $T_d$.

### B. Bit Framing

We transmit the data in small packets, or frames, each composed of a preamble, a payload, and a checksum.

**Preamble.** The preamble consists of a sequence of eight alternating bits ('10101010'), and it helps the receiver determine the properties of the channel, such as $T_{on}$ and $T_{off}$. In addition, the preamble header allows the receiver to synchronize with the beginning of a transmission and calibrate other parameters, such as the LEDs' location and the IR levels. **Payload.** The payload is the raw data to be transmitted. In our case, we choose 256 bits as the payload size. **CRC.** For error detection, a 16 bit CRC (cyclic redundancy check) value is added to the end of the frame. The receiver calculates the CRC for the received payload, and if it differs from the received CRC, an error is detected.

### C. Video Processing and Data Decoding

For video processing we use OpenCV [40], which is an open-source computer vision library that focuses on real-time video processing for academic and commercial use. Our program obtains the video stream recorded from the camera and calculates the IR LEDs' timings and amplitude. To detect and enumerate IR flickering, we use the following fundamental approach used in LED based communication [41] [42]. For each LED in the frame, we calculate the brightness function $p_n(x, y)$, where $x, y$ are the coordinates of a pixel in the image, and $n$ is a frame number. The algorithm calculates the intensity vector $S_{x,y}(p_0, p_1, ..., p_N)$ which describes the change of the illumination over time. The brightness of the LED is calculated as the quantized level of light intensity in the image at a point in the two-dimensional space. The algorithm determines the on and off brightness threshold values using the temporal mean of the sampled signals. Based on the intensity vector and threshold values, we demodulate the signals encoded in the video.

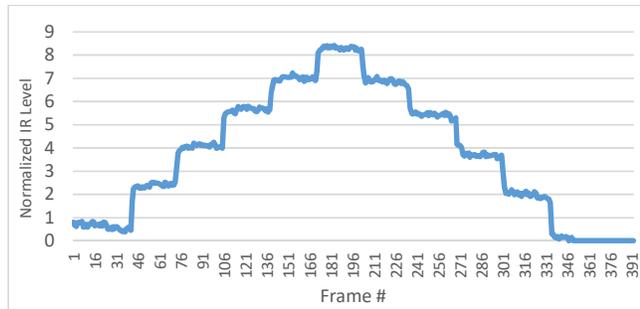

**Figure 10.** ASK modulation decoded from a video stream containing IR signals.

Figure 10 presents the output of the video processing algorithm on a video stream that contains IR signals. As can be seen in the figure, the data is encoded with ASK modulation with five amplitudes.

## VI. SIGNAL GENERATION

In this section we discuss the implementation of the exfiltration and infiltration channels.

### A. Exfiltration (via camera IR LEDs)

The IR LEDs in surveillance cameras can be controlled by the appropriate API provided by its firmware. In the most basic way, the state of the IR LEDs can be adjusted from within the camera's Web interface. Figure 11 shows the Web interface provided for the Sony SNC surveillance cameras [43]. The user can set the night vision to manual/automatic mode, in order to turn the IR LEDs on and off and set the level of the IR illumination.

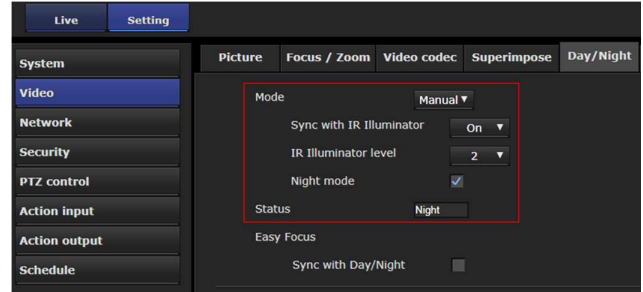

**Figure 11.** The Web interface for the Sony SNC-EM602RC camera, with the IR LEDs' control.

A malware needs to control the IR LED programmatically. One option for malware is to change the state of the IR LEDs by sending a POST request directly to the camera's HTTP server. The request will configure the IR LEDs' states according to the form fields provided by the Web interface. A more generic option is to use the control API provided with the camera SDK. The API allows applications to control the IR illumination and night vision settings from a user program. For example, the Sony SNC-EM602RC implements the LED IR control in the dynamic link library `snccgiw_dotnet.dll` within the `Sony.SNC.CGIWrapper` class. The two main functions that control the IR LEDs are `SetDayNightStatus` and `SetIrIlluminatorMaxStrength`. Invoking these functions causes an HTTP POST request to be sent to the camera's IP. Figure 12 shows the HTTP POST request sent to the camera (at `192.168.3.14`) over the network, when the IR illumination level is changed through the camera's SDK. In this example the `IRLed MaxStrength` parameter is set to '1'.

![Figure 12 showing HTTP request packet data]

**Figure 12. The HTTP request for changing the IR intensity of the Sony camera.**

For testing and evaluation, we implemented a program which encodes a binary file and transmits it via the IR LEDs. The program receives the camera's IP, the encoding along with the IR intensities' (amplitudes) timing parameters $T_{on}$ and $T_{off}$, and the binary file to transmit. The program uses the camera's SDK to control the IR LEDs and transmit the file according to the selected encoding scheme and parameters.

B. *Infiltration (via external IR LEDs)*

Since surveillance cameras can receive light in the IR wavelength, it is possible to deliver data into the organization through the video stream recorded by the surveillance cameras, using covert IR signals. For testing and evaluation, we implemented a system which can encode binary data over IR signals. We used a 12 volt IR projector consisting of 140 (10x14) IR LEDs, with an IR illumination distance of 130 meters. The LED switching is controlled by a Raspberry Pi via GPIO pin 40, which is connected to a MOSFET transistor (30N06LE) that in opened with three volts. We control the IR LEDs with a Python script which receives the encoding scheme along with the timing parameters $T_{on}$ and $T_{off}$, and the binary file to transmit. The setup is presented in Figure 13.

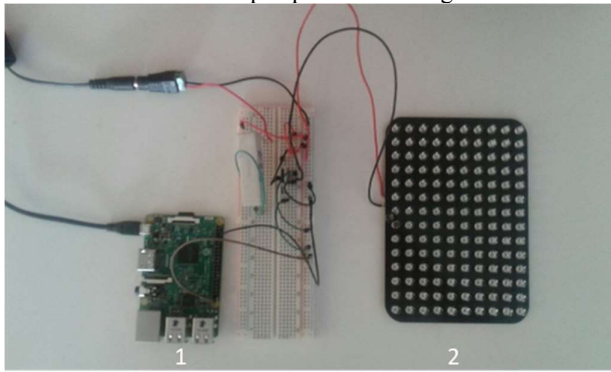

**Figure 13. The setup for the IR signal generation. A Raspberry Pi (1) controls the IR projector (2) which consists of 140 LEDs.**

## VII. EVALUATION & ANALYSIS

We evaluated and analyzed the exfiltration and infiltration covert channels to examine their maximum bit rate and distance. For the exfiltration scenario we tested security cameras and used their IR LEDs to leak data. We checked how the emitted IR signals were received from various types of video cameras. For the infiltration scenario we used the setup presented in Figure 13 and used it to transmit modulated binary data. For the transmission we used the professional surveillance camera Sony SNC-EM602RC [43]. This camera is an indoor and outdoor camera that includes built-in IR LEDs, allowing the viewing of objects up to 30 meters away in complete darkness. This camera includes 20 IR LEDs with IR wavelength of 850 nm. The camera has five different intensities of IR illumination.

A. *Exfiltration*

We evaluated how the IR signals sent from the security camera are received by five types of camera receivers. These cameras represent different scenarios which an attacker can use to record the exfiltrated data.

*1) Bit Rate*

There are two main factors limiting the bandwidth in a case of exfiltration: the IR LED transition time and the frame rate measured in frames per second (FPS) of the receiving camera.

- **IR LED transition time.** Our experiments on the Sony camera show that it takes about 50 ms to change the intensity of the IR LEDs from one level to another. This limitation may be a result of the LED hardware controller as well as the firmware implementation. This effectively limits the exfiltration bit rate to $20*\log(n)$ bit/sec, where $n$ is the number of levels used in the modulation. Note that in the case of OOK, $n = 2$. Note that this transition time may vary between different camera models and vendors.

- **The receiver FPS.** The bit rate is also bounded by the frame rate of the receiving camera. We identify 2-3 frames as the minimal number of frames required to detect the presence of IR signals, e.g., the presence of illuminating IR LEDs on two sequential frames is the indication of a signal.

Table 6 summarizes the maximal bit rate for the exfiltration channel for OOK and ASK modulation. The SNC-EM602RC camera was used for transmission, and five types of video cameras were used for the reception. For simplicity we used four levels of the IR illumination in the ASK modulation (out of five possible levels). The exfiltration channel is limited to 20 bit/sec with OOK and 40 bit/sec for ASK modulation. This makes the exfiltration channel more suitable for the transmission of a small amount of data such as passwords, PINs, encryption keys, and so on.

**Table 6. Max bit rate for exfiltration with OOK and ASK modulations using different camera receivers**

| Tested Camera/Sensor | FPS | Max bit rate (OOK) | Max bit rate (ASK 4 levels) |
|---|---|---|---|
| Smartphone camera (Samsung Galaxy J7) | 30 - 60 | 15-20 bit/sec | 30-40 bit/sec |
| Extreme camera (GoPro HERO5) | 60 - 240 | 20 bit/sec | 40 bit/sec |
| Entry-level DSLR (Nikon D7100) | 60 | 20 bit/sec | 30 bit/sec |
| Webcam (HD) (Microsoft LifeCam) | 30 | 15 bit/sec | 30 bit/sec |
| High-end security camera (Sony SNC-EB600) | 30 | 15 bit/sec | 30 bit/sec |

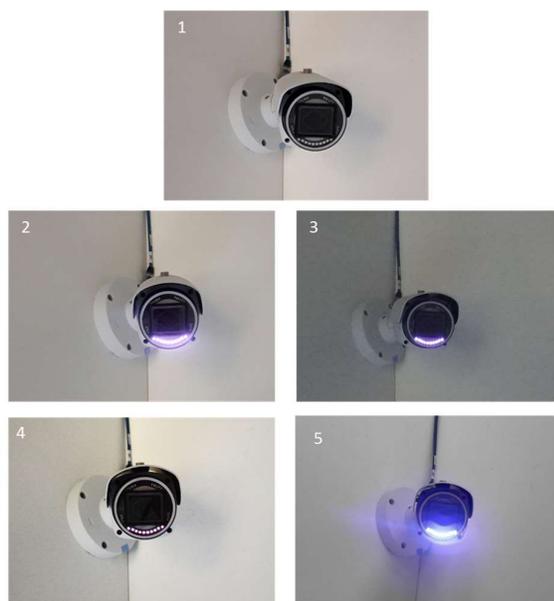

**Figure 14. The transmitting surveillance camera, as it seen by a human observer (1) and different camera receivers (2-5).**

Figure 14 shows the transmitting camera as it seen by (1) a human observer (the IR light is invisible), (2) a smartphone camera, (3) a professional video camera, (4) an extreme camera, and (5) an HD webcam.

*2) Stealth*

Since we modulate the data by changing the IR illumination, the change in the IR level is reflected in the video recorded by the security camera. It is possible that a person carefully watching the *recorded* video will notice the abnormal behavior in terms of the image brightness and reveal the covert channel. However, we found that shifting in the levels of IR illumination are difficult for a human observer to detect. This is mainly due to the brightness stabilizing techniques applied to the video stream. The following figures illustrate the transmission of `'01'` using level 0 (Figure 15) and level 1 (Figure 16) IR levels, as transmitted and recorded by the Sony SNC-EM602RC camera.

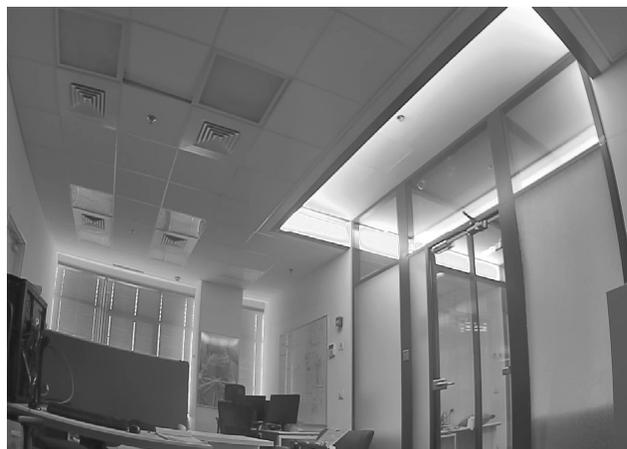

**Figure 15. Transmission of '0' using IR level 0 as viewed from the transmitting camera.**

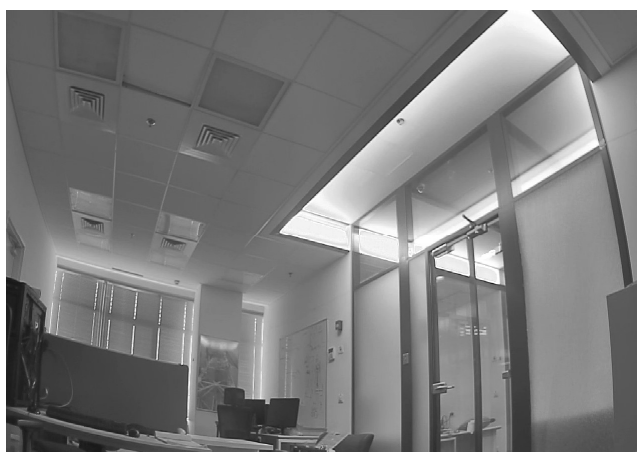

**Figure 16. Transmission of '0' using IR level 1 as viewed from the transmitting camera.**

**Table 7. The RGB values of the image in Figure 15**

|  | Avg | Med | Min | Max |
|---|---|---|---|---|
| **RGB:R** | 121 | 112 | 29 | 249 |
| **RGB:G** | 121 | 112 | 29 | 249 |
| **RGB:B** | 121 | 112 | 29 | 249 |

**Table 8. The RGB values of the image in Figure 16**

|  | Avg | Med | Min | Max |
|---|---|---|---|---|
| **RGB:R** | 122 | 114 | 32 | 246 |
| **RGB:G** | 122 | 114 | 32 | 246 |
| **RGB:B** | 122 | 114 | 32 | 246 |

Table 7 and Table 8 shows the average, median, minimum, and maximum values of the RGB values in Figure 15 and Figure 16, respectively. As shown in the tables, the brightness differences between the two images are negligible and difficult for a human observer to notice.

## B. Infiltration

LED (including IR LEDs) to camera communication has been studied and analyzed in prior research [44] [45]. The maximum bit rate and distance are mainly dependent upon the number of transmitting LEDs and the quality of the receiving camera. In this section we only examine the characteristic unique to the surveillance camera as a receiver. For more details on visible light and LED to camera communication, we refer the interested reader to [44] [45]. For testing and evaluation we generate IR using the transmitter described in Section VI. For the receiver we used two high-end Sony security cameras as well as HD webcam.

### 1) Bit Rate

The FPS of the security camera is the main factor limiting the bit rate per LED. However, in the case of infiltration, the attacker may use a number of LEDs to increase the bit rate. Table 9 compares the maximal bit rate for various configurations of transmitters and receivers.

**Table 9. Maximal bit rate for infiltration using one and eight IR LEDs and different cameras**

| Tested Camera/Sensor | FPS | Max bit rate per IR LED | Max bit rate using eight IR LEDs |
|---|---|---|---|
| High-end security camera (Sony SNCEB600) | 30 | 15 bit/sec | 120 bit/sec |
| High-end security camera (Sony SNCEB602R) | 30 | 15 bit/sec | 120 bit/sec |
| Webcam (HD) (Microsoft LifeCam) | 30 | 15 bit/sec | 120 bit/sec |

### 2) Stealth

The IR signals sent to the camera can be seen in the video stream as a white light. If the IR signals are strong, they may be detected by a person who observes the surveillance camera. In order to remain stealth, the IR signals must remain under certain thresholds so that they will be absorbed in the background of the images on the surveillance camera. For example, the attacker may use a small number of LEDs for transmission and only use them for a brief amount of time.

## C. Distances

One property of the covert channel is the communication range. In this section we discuss the communication ranges for the LOS and NLOS communication.

### 1) LOS Communication

The optical channel model is illustrated in Figure 17. The signal source is an array of IR LEDs on a surveillance camera that works in the NIR range. The typical receiver includes an appropriate optical filter to reduce the influence of the artificial lighting and illumination from the sun. Afterwards, the signal light is concentrated toward a photodetector (PD) by an optical lens system. The angular distribution of such lighting LED is commonly modeled by a generalized Lambertian radiant intensity model (measured in [sr$^{-1}$]) of the form

$$R(\phi) = \frac{m+1}{2\pi} \cos^m(\phi), \quad (1)$$

where $\phi$ is the irradiance angle and

$$m = -\frac{1}{\log_2[\cos(\Phi_{1/2})]} \quad (2)$$

is the Lambertian order with a semiangle of half power $\Phi_{1/2}$. The received power is calculated by

$$P_r = P_t R(\phi) \Omega L \cos(\varphi), \quad (3)$$

where $\Omega$ is the solid angle, $P_t$ is the LED power, $L$ is the optical system loss factor, and $\varphi$ is the axial alignment of the detector. The approximated solid angle (measured in [sr]) is calculated by

$$\Omega = \frac{A_l}{d^2} \quad (4)$$

where $A_l = \pi R_l^2$ is the area of the outer concentration lens with radius $R_l$, $d$ is the distance between the LEDs, and $A_l \ll d^2$ is assumed. For simplicity, multiple LED arrays inside a camera are modeled by a single powerful LED. The received signal (measured in A) is calculated by

$$i_r = \eta P_r + i_n, \quad (5)$$

where $\eta$ is the photodetector responsivity (measured in A/W), and $i_n$ is the electrical noise of the receiver. This noise is typically modeled by zero-mean Gaussian distribution with standard deviation $\sigma_n$. The resulting condition for a sufficiently low bit error rate is $\eta P_r \geq 10\sigma_n$. Finally, the maximum distance is calculated by

$$d \leq R_l \cos^{m/2}(\phi) \sqrt{\frac{\eta P_t L \cos(\varphi)(m+1)}{20\sigma_n}} \quad (6)$$

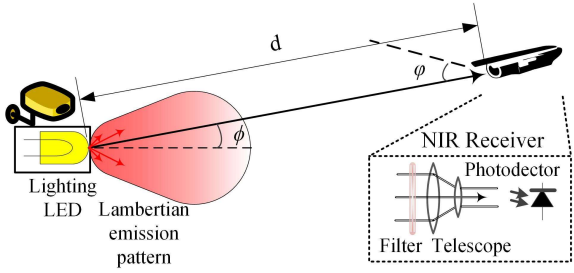

**Figure 17. Schematic illustration of an optical channel between the surveillance camera and a designated receiver.**

In a case of exfiltration, the signal source of the transmitter is the surveillance camera's LEDs. Since different IR levels are used for modulation, the power difference between these levels ($\Delta P_t$) must be applied to Eq. (1). The analysis of the communication channel for the typical values in Table 10 shows a communication distance of about 160 meters. Significant axial misalignment (higher $\phi$ angle) may significantly reduce this distance, while appropriate optical lenses may significantly increase this distance up to an order of magnitude.

**Table 10. Evaluation of the effective distance**

| Symbol | Value | Typical Range |
|---|---|---|
| $\Phi_{1/2}$ | 25° | 15°-30° |
| $\varphi$ | 5° | 0°-10° |
| $L$ | 0.8 | 0.75-0.95 |
| $R_l$ | 2.5cm | 1.5mm-5cm |
| $\eta$ | 0.5 | 0.5-0.8 |
| $\phi$ | 25° | |
| $\sigma_n$ | $10^{-8}$ A | $10^{-6}$-$10^{-11}$ A |
| $\Delta P_t$ | 4 W | 1-20 W |

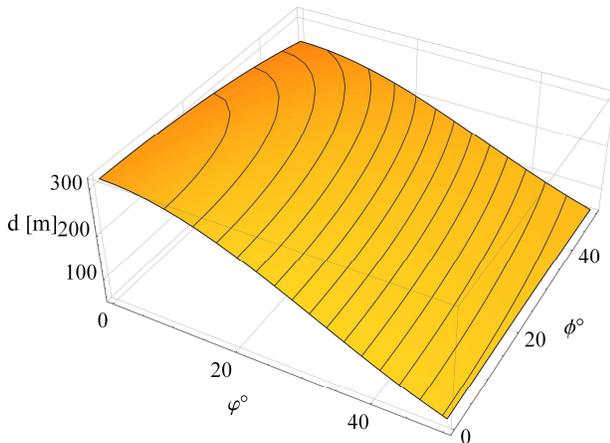

**Figure 18. The effective distance as the function of the irradiance angle, $\phi$, the axial detector angle, $\varphi$.**

The effective distance as the function of the irradiance angle, $\phi$, the axial detector angle, $\varphi$, and the parameters from Table 10.

The results show that both camera position and careful receiver alignment have notable influence on the communication distance.

In a case of infiltration, theoretical analysis of the maximum distance highly depends on the parameters of the transmitter, such as the power of the light source. It also depends on the surveillance camera's technical specifications, such as the resolution, focal lens, and more. However, previous works on the topic of LED to camera communication show that such communication is possible for distances of tens to hundreds of meters between the transmitter and the camera receiver. The rigorous analysis of LED to camera communication is out of the scope of this paper, and we refer the interested reader to several works in this field [46] [47] [48].

*2) NLOS Communication*

The operation principle of "night-vision" lighting is based on the diffuse back-reflection of light from the area of interest. However, this diffuse-reflected light may be used for NLOS exfiltration and infiltration.

The common model that is based on Lambertian cosine reflection law is given by [49]

$$P_r = P_t \int_{area} R_l^2 \rho \frac{m+1}{4\pi} \cos^m(\phi) \cos(\alpha) \times \cos(\beta) \frac{1}{D_1^2} \frac{1}{D_2^2} \cos(\varphi) dA, \quad (7)$$

where $\alpha$ is the incidence angle, $\beta$ is the reflection angle, $D_1$ and $D_2$ are distances between the LED and the reflecting surface and between the surface and the detector, respectively (outlined in Figure 19), and $\rho$ is the reflection coefficient. The integral is calculated over the whole illuminated area. It is important to note that the decrease in signal strength is proportional to the fourth power of the distance ($\sim d^{-4}$).

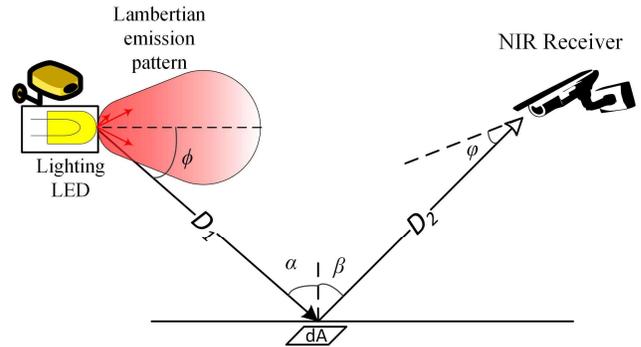

**Figure 19. Schematic illustration of NLOS communication between the surveillance camera and a designated receiver.**

With high-end surveillance cameras, the effective exfiltration distance for NLOS communication is between 0 and 30 meters. This distance may be improved by using high quality optical sensors and lenses. Communication for greater distances is also available in some special scenarios, e.g., when the transmitting camera faces a wall or other types of reflective objects ($\alpha$=0).

Table 11 summarizes the feasible ranges for LOS and NLOS communication for the exfiltration and infiltration channels.

**Table 11. Effective distances of the LOS and NLOS communication**

|  | Line-of-sight | Non line-of-sight |
|---|---|---|
| Exfiltration | Tens to hundreds of meters | Tens of meters |
| Infiltration | Tens of meters to Kilometers | Tens of meters |

## VIII. COUNTERMEASURES

General preventive countermeasures against optical emanation may include policies aimed to restrict the (optical) visibility of the emanated equipment. For example standards such as the NATO SDIP-27 (levels A/B/C) and SDIP-28 define classified zones for electronic equipment that can emanate signals [50] [51] [52] [53]. In the context of our attack, an organization may place surveillance cameras in restricted zones that are optically inaccessible to attackers. This type of countermeasure is less relevant to surveillance cameras that are intentionally pointing at public areas (e.g., at the street). For indoor cameras, it is possible to install an opaque window film that prevents optical visibility from outside the building [54]; note that this type of countermeasure doesn't protect against insider attacks or cameras located within the building.

### A. Exfiltration

Disabling the IR LEDs in the surveillance cameras may prevent the exfiltration channel presented in this paper. This strategy has two main drawbacks. First, disabling the IR LEDs effectively disables the night vision capability, which is an essential function of surveillance cameras. Second, as long disabling the IR LEDs is done at the software level (e.g., by changing the default settings), malware can re-enable it again. Even if disabling the IR LEDs is done at the firmware level, the surveillance camera itself may be compromised by malware (e.g., firmware level rootkit [55]) that can override these settings [56] [26] [57] [58]. A less elegant countermeasure against exfiltration attacks is to physically disconnect the camera's IR LEDs, or cover them with black tape which blocks the optical emanation [19]. Like the previous solution, such countermeasures effectively disable the night vision capability of the camera and decrease its usability. To prevent the exfiltration channel it is possible to implement a security component within the camera firmware. There are two types of firmware level security mechanisms that protect against the covert channels presented in this paper. The first approach is to add a minimal timeout between alterations to the status of the IR LEDs. Effectively, this will bound the potential bandwidth of the covert channel. The second approach - also known as signal jamming - is to invoke random IR blinking in the camera, so that the signals from the malicious code will get mixed up with the generated noise. Implementation of such a timeout mechanism or noise generator within the camera requires the involvement of the camera manufacturer and may also affect the usability of the IR LEDs.

### B. Infiltration

The infiltration channel can be prevented by adding an IR filter to the surveillance camera. That way, all IR signals will not pass through the recorded video. Although this solution is hermetic, it effectively eliminates the night vision capability of the camera, which is based on IR light. The infiltration channel can be detected mainly by analyzing the recorded video stream, searching for irregular LED activities. Note that such a solution must be done by performing offline video analysis. An LED detection solution may also suffer from a high rate of false positives in certain environments, e.g., it may raise false alerts for road lights and car lights.

### C. Detecting Camera-Related Malware

Technological countermeasures may include the detection of the presence of malware that controls the camera's IR LEDs or monitors the camera's input. The detection can take place at the host or network level. Detection at the host level involves the monitoring of irregular access to the network camera configurations' APIs. For example, frequent access to the API may indicates that the IR LEDs are being manipulated for a malicious purpose. However, as with any other host-based detection, this approach can be evaded by malware which also resides on the host. Similarly, detection can be done at the network level, by monitoring the network traffic from hosts in the network to the surveillance cameras. In this case, it is possible to detect irregular access to the camera or identify the messages aimed at configuring the IR LEDs or reading the camera's input. Such a network-based detection requires deep packet inspection which might not always be possible (e.g., in the case of encrypted traffic).

The countermeasures are summarized in Table 12.

**Table 12. List of countermeasures**

| Type | Countermeasure | Limitations | Relevancy |
|---|---|---|---|
| Prevention | Zoning, camera banning, and area restriction | Insider threats (e.g., indoor cameras and malicious insiders) | Infiltration/ exfiltration |
| Prevention | LEDs covering / disconnecting | Degradation of the night vision functionality and the user experience | Exfiltration |
| Prevention | Window shielding | High cost | Infiltration/ exfiltration |
| Prevention (software / firmware) | Timeout mechanisms, disabling IR control | Requires OEM support, can | Exfiltration |

| | | be evaded by malware | |
| --- | --- | --- | --- |
| Prevention | Signal jamming | Degradation of night vision functionality | Infiltration/ exfiltration |
| Prevention | Adding IR filter | Degradation of night vision capability | Infiltration |
| Detection | LED activity monitoring (external) | Price | Infiltration/ exfiltration |
| Malware Detection (network) | Suspicious traffic detection (LED control) | Not relevant if traffic is encrypted | Exfiltration |
| Malware Detection (host) | Irregular access to camera's API | Can be evaded by malware | Infiltration/ exfiltration |
| Detection (video) | Detect IR activity from the recorded video | High false positive rate | Infiltration/ exfiltration |

IX. CONCLUSION

Infrared light is invisible to humans but can be optically recorded by many types of cameras. In this paper we show how attackers can exploit indoor and outdoor surveillance cameras for data exfiltration and infiltration via IR illumination. In the exfiltration scenario, malware accesses the surveillance camera in the local network and generates covert IR signals by controlling the IR LEDs. Binary data is modulated, encoded, and transmitted over these signals. An attacker from a distance away with a line of sight to the surveillance camera can receive the IR signals and decode the binary information. In the infiltration scenario, covert IR signals are generated by a remote attacker with IR LEDs. These signals are then received by the surveillance camera and intercepted by malware within the network. We describe the attack model and provide scientific background about this optical covert channel, such as line-of-sight (LOS) and non-line-of-sight (NLOS) communication. We present data modulation schemas and a basic transmission protocol. We implement a malware prototype, evaluate it with different types of cameras and discuss preventive and defensive countermeasures. Our evaluation shows that an attacker can use IR and surveillance cameras to communicate over the air-gap to a distance of tens to hundreds of meters away. We demonstrate how data can be leaked from the network at a bit rate of 20 bit/sec (per camera) and be delivered to the network at bit rate of more than 100 bit/sec (per camera).